\documentclass[12pt]{article}
\usepackage{a4wide}
\usepackage{amssymb}
\usepackage{mathrsfs}
\usepackage{comment}
\usepackage{tensor}
\usepackage{amsmath}
\usepackage{graphicx}

\usepackage[dvipsnames]{xcolor}

\begin{document}
{\renewcommand{\thefootnote}{\fnsymbol{footnote}}
\hfill  \\
\medskip
\begin{center}
{\LARGE Singularities in loop quantum cosmology}\\
\vspace{1.5em}

Martin Bojowald,$^1$\footnote{e-mail address: {\tt bojowald@psu.edu}}
Manuel D\'{\i}az,$^2$\footnote{e-mail
address: {\tt manueldiaz@umass.edu}}
and
Erick I. Duque$^1$\footnote{e-mail address: {\tt eqd5272@psu.edu}}

\vspace{1em}
$^1$ Institute for Gravitation and the Cosmos,
The Pennsylvania State
University,\\
104 Davey Lab, University Park, PA 16802, USA\\
\vspace{0.5em}

$^2$  Amherst Center for Fundamental Interactions,\\ Department of Physics,
University of Massachusetts Amherst,\\
426 Lederle Graduate Research Tower,
Amherst, MA 01003 USA\\

\vspace{1.5em}
\end{center}
}

\setcounter{footnote}{0}

\begin{abstract}
  Quantum effects are expected to modify the cosmological dynamics of the
  early universe while maintaining some (potentially discrete) notion of
  space-time structure. In one approach, loop quantum cosmology, current
  models are shown here to either be incompatible with a consistent space-time
  structure, or to have physical singularities. The latter happens in spite of
  a non-zero scale factor in the isotropic background dynamics. A new
  effective Friedmann equation shows that a bounce is obtained at sub-Planckian
  densities, preceded by a physical singularity at infinite scale factor that
  resembles a time-reversed big rip. The entire phase is accompanied by rapid
  changes of the Hubble radius.  In addition, a new version of perturbative
  inhomogeneity in loop quantum cosmology is introduced that maintains a
  consistent space-time structure and has a non-singular background dynamics.
\end{abstract}

\section{Introduction}

Physical singularities in solutions of general relativity impose limitations
on this theory that should be resolved by including new features in the
underlying equations of motion. A general expectation is that quantum
properties not only of matter but also of space-time should play an important
role in these extreme regimes of immense density and curvature. But
formulating a well-defined quantum theory of space-time is complicated by
several long-standing conceptual and mathematical problems, one of them being
a reconciliation of discrete quantum aspects with continuous symmetries, such
as Lorentz boosts, conventionally used to define the structure of continuous
space-time.

In order to explore potential implications of quantum theory applied to
space-time, it has become popular to work in a reduced context of homogeneous
and isotropic cosmological models. Homogeneity implies the existence of a
preferred frame that can be used to restrict non-Abelian Lorentz
transformations to a simple reparameterization symmetry of the time
coordinate. Many mathematical challenges that would have to be faced for a
full quantum description of general relativity then disappear, and dynamical
implications can be studied by a standard analysis of modified Friedmann
equations. The resulting solutions, interpreted as space-times, can be
used as backgrounds for other fields or for perturbative inhomogeneity, making
it possible to derive potential implications for early-universe cosmology.

As an example, loop quantum cosmology \cite{LivRev} is often presented as
replacing the classical flat-space Friedmann equation
\begin{equation}
  {\cal H}^2=\frac{8\pi G}{3}\rho
\end{equation}
for the Hubble parameter ${\cal H}=\dot{a}/a$ in terms of the scale factor $a(\tau)$
as a function of proper time, using the energy density $\rho$ of matter, with
the modified equation \cite{GenericBounce,AmbigConstr,APSII}
\begin{equation} \label{FriedmannMod}
  {\cal H}^2=\frac{8\pi G}{3} \rho\left(1-\frac{\rho}{\rho_{\rm Q}}\right)
\end{equation}
characterized by a quantum density $\rho_{\rm Q}$. In the simplest cases,
$\rho_{\rm Q}$ is constant and of Planckian size. The modified Friedmann
equation then implies a vanishing $\dot{a}$ when the density reaches
$\rho=\rho_{\rm Q}$, which can be shown to imply a local minimum by using a
corresponding modified version of the Raychaudhuri equation. In this way, the
classical big-bang singularity may be replaced by a bounce.

Assessing the reliability of this statement requires an analysis of how the
modified Friedmann equation is derived. Loop quantum cosmology is based on the
canonical formulation of general relativity, which in the isotropic context
can be stated easily by viewing the Hubble parameter ${\cal H}=-4\pi G p_q$ as
proportional to the momentum $p_q$ of the volume $q=a^3$. The Friedmann equation
then takes the form of a Hamiltonian constraint,
\begin{equation}
C=  - 6\pi G qp_q^2 + q\rho=0\,.
\end{equation}
The expression $C$ has units of energy and can be used as a Hamiltonian,
implying Hamilton's equations
\begin{equation}
  \dot{q}=\frac{\partial C}{\partial p_q}= -12\pi Gqp_q=3a^2\dot{a}\,,
\end{equation}
consistent with the definition of $q$, and
\begin{equation}
  \dot{p}_q=-\frac{\partial C}{\partial q}= 6\pi G p_q^2-\frac{\partial
    (q\rho)}{\partial q}
\end{equation}
which, using $q\rho=E$ the matter energy, is equivalent to the Raychaudhuri equation
\begin{equation}
  \dot{{\cal H}}=-\frac{3}{2}{\cal H}^2-4\pi G P
\end{equation}
with the matter pressure $P=-\partial E/\partial V$.

Following general constructions in loop quantum gravity
\cite{ALRev,Rov:Loops,ThomasRev}, loop quantum cosmology proposes that the
polynomial dependence of $C$ on $p_q$ may be modified to become periodic,
replacing $p_q$ with $\sin(\ell p_q)/\ell$ with a length parameter $\ell$. The
classical equations are obtained in the limit $\ell\to0$, while a Planckian
$\ell$ leads to the modified Friedmann equation (\ref{FriedmannMod}) with a
Planckian $\rho_{\rm Q}$: Defining
\begin{equation} \label{Cl}
  C_{\ell}=-6\pi Gq \frac{\sin^2(\ell p_q)}{\ell^2}+q\rho=0
\end{equation}
now results in the first Hamilton's equation
\begin{equation} \label{dotqMod}
  \dot{q}=\frac{\partial C_{\ell}}{\partial p_q}=-12\pi Gq \frac{\sin(\ell
    p_q)\cos(\ell p_q)}{\ell} 
\end{equation}
such that $C_{\ell}=0$ implies
\begin{equation}
  \frac{\dot{q}^2}{9q^2} = 16\pi^2G^2\frac{\sin^2(\ell p_q)}{\ell^2}\left(1-\sin^2(\ell
    p_q)\right)
\end{equation}
or (\ref{FriedmannMod}) with
\begin{equation} \label{rhoQell}
  \rho_{\rm Q}= 6\pi G/\ell^2
\end{equation}
upon using (\ref{Cl}).

Equation~(\ref{dotqMod}) demonstrates that the classical relationship between
the momentum $p_q$ and the time derivative $\dot{q}$ is modified. The
relationship between the rate of change $\dot{q}/q$ and the canonical momentum
therefore changes too. Such a modification is not unexpected because the
momentum contains dynamical information, which may deviate from its classical
behavior in models of quantum dynamics. However, a second assumption is often
made in this context without further justifications: The relationship between
the observational Hubble parameter ${\cal H}$ and the rate of change
$\dot{q}/q$ of the canonical variable $q$ is assumed to remain unchanged,
given by ${\cal H}=\frac{1}{3}\dot{q}/q$ as in the classical
theory. Maintaining this relationship requires justifications because there is
in general no reason for a generic canonical variable $q$ to determine the
spatial expansion rate. Even though $q$ has a specific relationship with the
spatial volume in the classical theory, this property is not guaranteed to
hold after the theory has been modified by using $C_{\ell}$ instead of $C$.

The classical definition of the Hubble parameter is not dynamical but rather defines a
measurable parameter for the expansion rate of the universe at a given
time. Instead of being dynamical, this quantity is geometrical because it
relies on the classical notion of space-time geometry, encoded in the
isotropic context for flat spatial slices by the line element
\begin{equation}
  {\rm d}s^2=-{\rm d}\tau^2+a(\tau)^2({\rm d}x^2+{\rm d}y^2+{\rm d}z^2)\,.
\end{equation}
The coefficients in a line element cannot be chosen in an arbitrary manner but
require covariance properties, in general expressed as the tensor
transformation law: The geometrical interpretation of
(Riemannian) space-time stipulates that ${\rm d}s^2$ is directly measurable as
a proper length or proper time of space-time intervals. As such, it must be
independent of the coordinate choice $(\tau,x,y,z)$ used to specify its
coefficients. Defining modified dynamics in terms of the Hamiltonian
constraint $C_{\ell}$ makes sure that the time coordinate can still be
transformed after modifications: Setting the Hamiltonian equal to zero, rather
than a variable energy value, implies that there is no absolute time. However,
this one equation does not determine independent properties of spatial
coordinate transformations, which are required for a consistent interpretation
of $a(\tau)^3$ as the spatial volume (of a suitable region) at constant time $\tau$.

The continued use of $\frac{1}{3}\dot{q}/q$ as the Hubble parameter or of $q$ as the
spatial volume even after the dynamics has been modified is therefore based on
a hidden assumption about the underlying space-time geometry. This assumption
is hard to notice and impossible to test within strictly isotropic or
homogeneous models. But it is subject to non-trivial conditions if one tries
to embed such a model within a theory of perturbative inhomogeneity. Even
small inhomogeneity means that spatial transformations are non-trivial, just
as time transformations are non-trivial in a dynamical context. It is then
possible to investigate whether combinations of time and space coordinate
changes can be combined into a single space-time structure, locally described
by transformations of the Poincar\'e algebra.

Here, we present a suitable treatment which, generically, does turn out to
require modifications of the relationship between the isotropic canonical
variable $q$ and the spatial volume in loop quantum cosmology, just as the
dynamical behavior is modified by using $C_{\ell}$ instead of $C$. A brief
summary of the underlying conceptual and technical developments, given by
emergent modified gravity, is given in Section~\ref{s:Pert}. As a consequence,
shown in Section~\ref{s:Sing}, previously constructed models \cite{ScalarHol}
of perturbative inhomogeneity in loop quantum cosmology turn out to be
singular, even though they are formulated on a background following the
dynamics implied by (\ref{Cl}). We derive a new effective Friedmann equation
that captures this behavior, and analyze its immediate cosmological
implications such as a big-rip type initial singularity followed by a
bounce. In Section~\ref{s:NonSing}, we show that the same background dynamics
is compatible with different covariant forms of dynamical perturbative
inhomogeneity, some of which can have non-singular background
dynamics. However, they violate one of the main assumptions usually made in
models of loop quantum cosmology: the appearance of bounded periodic functions
of $p_q$ as coefficients in (\ref{Cl}) and its extensions by perturbative
inhomogeneity. Our findings therefore constitute a major paradigm shift for
modelling the effects of loop quantum gravity in a covariant manner.

\section{Perturbative inhomogeneity}
\label{s:Pert}

Different choices of the time coordinate are implemented canonically by
replacing the isotropic $C$ with an expression $H[N]=NC$ for an arbitrary
time-dependent lapse function $N$. This factor then modifies all on-shell time
derivatives in Hamilton's equations with $C=0$ such that ${\rm d}\tau$ is
equivalent to $N{\rm d}t$ with a new time coordinate $t$.

Similarly, a small spatial dependence can be introduced in the fields and in
coordinate changes by using a spatially integrated
\begin{equation} \label{ClassPertHamConst0}
H[N] =  \int_{\Sigma}\mathrm{d}^3x NC = H_1[\bar{N}] + H_2[\delta N] 
\end{equation}
where $\bar{N}$ now indicates the homogeneous background value and $\delta N$
is a small inhomogeneous perturbation in $N=\bar{N}+\delta N$. The new $C$
should be derived by a Taylor expansion of the full expression from general
relativity with respect to small inhomogeneity of all relevant fields. The
expansion must be made up to second order so as to produce first-order
Hamilton's equations. In vacuum with a cosmological constant, this procedure
leads to \cite{HamGaugePert,HamPerturb,ConstraintAlgebra}
\begin{equation}
\label{ClassPertHamConst}
H[\bar{N}] = \int \mathrm{d}^3x\bar{N}\left[ C^{(0)} + C^{(2)}\right]
    \quad,\quad 
    H[\delta N] =  \int \mathrm{d}^3x \delta N C^{(1)}
\end{equation}
where
\begin{eqnarray}\label{HamConstH0}
    C^{(0)} &=& -6\bar{k}^2 \sqrt{\bar p}
    + 2\Lambda \bar{p}^{3/2}
~,\\
\label{HamConstH1}
    C^{(1)} &=&
    - 4 \bar{k}\sqrt{\bar{p}} \delta^c_j\delta K_c^j
    -\frac{\bar{k}^2-\Lambda \bar{p}}{\sqrt{\bar{p}}} \delta_c^j\delta E^c_j
    +\frac{2}{\sqrt{\bar{p}}} \partial_c\partial^j\delta E^c_j
~,\\
\label{HamConstH2}
    C^{(2)} &=&
    \bar{p}^{3/2} \left(\frac{\delta K_c^j\delta K_d^k\delta^c_k\delta^d_j}{\bar{p}}
    - \frac{(\delta K_c^j\delta^c_j)^2}{\bar{p}}\right)
    - \frac{2\bar{k}}{\sqrt{\bar{p}}} \delta E^c_j\delta K_c^j
    \\
    && 
    - \frac{(\bar{k}^2 +\Lambda \bar{p})\sqrt{\bar{p}}}{2} \left(\frac{\delta_c^k\delta_d^j\delta E^c_j\delta E^d_k}{\bar{p}^2}
    - \frac{(\delta_c^j \delta E^c_j)^2}{2\bar{p}^2} \right)
    - \frac{\bar{p}^{3/2}}{2} \frac{\delta^{jk} (\partial_c\delta E^c_j)
       (\partial_d\delta E^d_k)}{\bar{p}^3}\,.\nonumber 
\end{eqnarray}
Here, we use canonical variables $\bar{p}=q^{2/3}$ and $\bar{k}=q^{1/3}p_q$ for
the background, and $\delta E^a_i$  and $\delta K_b^j$ for
inhomogeneity. (These choices simplify some of the spatial derivatives in the
constraint, which is convenient when integrating by parts in evaluations of
Poisson brackets.)

Classically, the space-time geometry can be reconstructed by computing the
spatial metric
\begin{equation}\label{qab}
    q_{ab} =\bar{p} \left( \delta_{a b} - \delta_{a c} \delta_{b d} \frac{\delta^{cj} \delta E^d_j
    + \delta^{dj} \delta E^c_j}{\bar{p}}
    + \delta_{a b} \frac{\delta_c^j\delta E^c_j}{\bar{p}} \right)\,,
\end{equation}
and $\delta K_b^j$ is related to the extrinsic curvature of a spacelike
hypersurface in a space-time with line element
\begin{equation} \label{ds}
  {\rm d}s^2= -N^2{\rm d}t^2+q_{ab}({\rm d}x^a+N^a{\rm d}t)({\rm d}x^b+N^b{\rm
    d}t)\,.
\end{equation}
While the canonical variables determine the spatial part of this line element,
the time components are given by the lapse function $N$ and
a shift vector $N^a=\delta N^a$ (of first order in the perturbed system) that
define the time-evolution vector field $t^{\mu}=Nn^{\mu}+N^as_a^{\mu}$
relative to the unit normal $n^{\mu}$ and a basis $s_a^{\mu}$ of embedded
spacelike hypersurfaces at constant $t$.
In addition to $H[N]$, there is a Hamiltonian generator of spatial transformations,
referred to as the diffeomorphism constraint
\begin{equation} \label{PertDiffConst}
D_c[\delta N^c] = \int_{\Sigma}\mathrm{d}^3x\delta N^c
\left(\bar{p}\partial_c(\delta^d_k \delta K^k_d)
-\bar{p}(\partial_k\delta K^k_c)- \bar{k} \partial_d (\delta_c^k\delta E^d_k)\right) \,.
\end{equation}
The complete evolution generator in Hamilton's equations for given lapse
function and shift vector is $H[N]+D_a[\delta N^c]$.

The relationship (\ref{qab}) follows from the derivation of the expanded
generators from general relativity, where the space-time metric was introduced
as the main basic field of an action principle. For modified canonical
theories, an important question is how a suitable $q_{ab}$ can be derived from
the canonical variables without assuming any correspondence with classical
general relativity. If some modified canonical theory is supposed to be
a candidate for an effective description of canonical quantum gravity, it must be possible to
derive a consistent geometrical interpretation from scratch, without relying
on classical relationships between metric components and canonical fields such
as $\bar{p}$ and $\delta E^a_i$.

A successful derivation relies on the seminal contributions to canonical
general relativity made in \cite{DiracHamGR,Katz,ADM,Regained} and has
recently been completed in the modified context by several contributions from
\cite{Absorb,EffLine,NonCovPol,SphSymmEff,Higher,HigherCov}.  A necessary
condition for the generators to correspond to coordinate transformations in a
space-time with line element (\ref{ds}) is that they have Poisson brackets of
the form
\begin{eqnarray}
    \{ D_a[N^a_1] , D_b [N^b_2] \} &=& - D_a [N_2^b \partial_b N_1^a-N_2^b \partial_b N_2^a]
    \ , \label{DD}
    \\
    \{ H [ N ] , D_a [ N^a]\} &=& - H [ N^b \partial_b N ]
    \ ,\label{HD} \\
    \{ H [ N_1 ] , H [ N_2 ] \} &=& - D_a [ q^{a b} ( N_2 \partial_b N_1 - N_1
                                    \partial_b N_2 )]\,. \label{HH} 
\end{eqnarray}
The first two equations can be seen to be canonical versions of the Lie
derivative along spatial vector fields $N^a$, which is extended to the time
dimension by the third equation. This last equation contains the inverse
$q^{ab}$ of the spatial metric in a space-time geometry in which the
corresponding coordinate changes can be applied. Symmetries of the canonical
theory therefore determine a compatible spatial metric.

In the context of perturbative inhomogeneity, this equation is turned into
\begin{eqnarray} \label{HHClassical}
\{H[N_1], H[N_2]\} &=&
\{H[\delta N_1],H[\bar{N}_2]\}+
\{H[\bar{N}_1],H[\delta N_2]\}\\
&=& D_a\left[\bar{q}^{a b} (\bar{N}_2 \partial_b\delta N_1-\bar{N}_1 \partial_b\delta N_2)\right] \,.\nonumber
\end{eqnarray}
Space-time symmetries therefore determine the background metric
$\bar{q}_{ab}$, which is sufficient for testing the assumption that $q$ remains
a valid spatial volume, and ${\cal H}=\frac{1}{3}\dot{q}/q$ a valid Hubble parameter
in a modified theory, such as loop quantum cosmology.

Following the perspective of effective field theory, all terms in a generic
Hamiltonian that are expected to have comparable magnitude while respecting
essential symmetries should be included. Comparable magnitude is usually
determined by derivative orders because higher energies correspond to stronger
variations in fields or wave functions.  In canonical form, temporal
variations are determined by equations of motion and do not appear directly in
Hamiltonians. Higher time derivatives instead correspond to new independent
momenta because an $n$th-order equation of motion implies that the field and
its first $n-1$ derivatives can be chosen freely in initial data. We should
therefore arrange terms by their spatial derivative order combined with the
number of independent momenta. Here, we work with the classical orders, which
means that we do not introduce additional perturbative momenta beyond $\delta
K_a^i$, and include all terms with up to two spatial derivatives in any
product. The perturbative nature then allows us to write the generic constraint
contributions as 
\begin{equation}
\label{HamConstH0-EMG}
    \tilde{C}^{(0)} = - 6 \sqrt{\bar p} {\cal K}^{(0)}
\end{equation}
for the background and
\begin{equation}
\label{HamConstH1-EMG}
    \tilde{C}^{(1)} =
    \sqrt{\bar{p}} \left[- 4 {\cal K}^{(1)}_1 \delta^c_j\delta K_c^j
    - {\cal K}^{(1)}_2 \frac{\delta_c^j\delta E^c_j}{\bar{p}}
    + 2 {\cal K}^{(1)}_3 \frac{\partial^j\partial_c\delta E^c_j}{\bar{p}} \right]
\end{equation}
as well as
\begin{eqnarray}  
\label{HamConstH2-EMG}
    \tilde{C}^{(2)} &=&
    \sqrt{\bar{p}} \Bigg[ {\cal K}^{(2)}_1 \delta K_c^j\delta K_d^k\delta^c_k\delta^d_j
    - {\cal K}^{(2)}_2 (\delta K_c^j\delta^c_j)^2
    - 2 {\cal K}^{(2)}_3 \delta K_c^j \frac{\delta E^c_j}{\bar{p}}
    \nonumber\\
    && \quad 
    - \frac{1}{2} \left({\cal K}^{(2)}_4 \frac{\delta_c^k\delta_d^j\delta E^c_j\delta E^d_k}{\bar{p}^2}
    - {\cal K}^{(2)}_5 \frac{(\delta_c^j \delta E^c_j)^2}{2\bar{p}^2} \right)
    - \frac{{\cal K}^{(2)}_6}{2} \frac{\delta^{jk} (\partial_c\delta E^c_j)
       (\partial_d\delta E^d_k)}{\bar{p}^2} \Bigg] 
\end{eqnarray}
for perturbations. The ten functions ${\cal K}^{(0)}$, ${\cal K}_I^{(1)}$
and ${\cal K}_I^{(2)}$ are free functions of $\bar{p}$ and $\bar{k}$ at this
stage. Up to integrations by parts, all second-order terms are included.

We combine these expressions with the classical diffeormorphism constraint in
order to maintain standard spatial coordinate changes. Several conditions are
then imposed, which are direct implications of the symmetry requirements but
are quite involved in their explicit canonical realization: (i) We require
that brackets of the form (\ref{DD})--(\ref{HH}) are satisfied, possibly with a
modified phase-space function $\bar{\tilde{q}}^{ab}$ in (\ref{HHClassical})
that does not obey the classical relationship with $|p|^{-1}\delta^{ab}$. (ii)
The new function $\bar{\tilde{q}}^{ab}$ or a perturbative extension by some
$\delta \tilde{q}^{ab}$, as a phase-space function, is subject to Hamilton's
equations generated by $\tilde{H}$ and $D$. These transformations must be
equivalent to infinitesimal coordinate changes of an inverse spatial metric
when the constraints $\tilde{H}[N]=0$ and $D[N^a]=0$ are imposed.

The resulting conditions on the ${\cal K}$-functions, derived in
\cite{EMGCosmoK}, can be stated rather compactly:
\begin{eqnarray}
    \label{eq:K^2_2=K^2_1 - simp}
    {\cal K}^{(2)}_2&=&{\cal K}^{(2)}_1
    \,,\\
    \label{eq:dK^0/dk - simp}
    {\cal K}^{(2)}_3 &=&
    \frac{\partial {\cal K}^{(0)}}{\partial \bar{k}}
    - \bar{k} {\cal K}^{(2)}_1
    \,,\\
    \label{eq:dK^0/dp - simp}
    {\cal K}^{(2)}_4
    &=& 2 \bar{k} {\cal K}^{(2)}_3
    -{\cal K}^{(0)}
    - 2 \bar{p} \frac{\partial {\cal K}^{(0)}}{\partial \bar{p}}
    \,,\\
    \label{eq:K^2_5=K^2_4 - simp}
    {\cal K}^{(2)}_5&=&{\cal K}^{(2)}_4
    \,,\\
     \label{eq:HH=D - simp}
    2 {\cal K}^{(1)}_1 {\cal K}^{(2)}_6 &=&
    \frac{\partial {\cal K}^{(0)}}{\partial \bar{k}} \left({\cal K}^{(1)}_3
                                            \frac{\partial {\cal
                                            K}^{(0)}}{\partial \bar{k}} 
    + 2 \bar{p} \frac{\partial {\cal K}^{(1)}_3}{\partial \bar{p}}\right)
    - \frac{\partial {\cal K}^{(1)}_3}{\partial \bar{k}} \left({\cal K}^{(0)}
    + 2 \bar{p} \frac{\partial {\cal K}^{(0)}}{\partial \bar{p}}\right)
    \,,\\
  \label{eq:HH cond E - simp}
     -3 {\cal K}^{(0)} {\cal K}^{(2)}_1
  &=& 2 \bar{p} \frac{\partial {\cal K}^{(1)}_1}{\partial \bar{p}} \frac{\partial {\cal K}^{(0)}}{\partial \bar{k}}
    - \frac{\partial {\cal K}^{(1)}_1}{\partial \bar{k}} \left({\cal K}^{(0)}
    + 2 \bar{p} \frac{\partial {\cal K}^{(0)}}{\partial \bar{p}}\right)
    - {\cal K}^{(1)}_1 \frac{\partial {\cal K}^{(0)}}{\partial \bar{k}}
      \,,\\
    \label{eq:K12 - simp}
    {\cal K}^{(1)}_2 &=& 3 {\cal K}^{(0)}
    -2 \bar{k} {\cal K}^{(1)}_1
    \,.
\end{eqnarray}
The first six conditions directly eliminate the second-order modification
functions ${\cal K}_I^{(2)}$ (since ${\cal K}^{(0)}$ and ${\cal K}_1^{(1)}$
cannot be identically zero in order to observe the correct classical
limit). The first-order function ${\cal K}_2^{(1)}$ is fixed by the last
condition. It is therefore possible to assume some background modification
${\cal K}^{(0)}$ as a function of $\bar{p}$ and $\bar{k}$ and two first-order
terms, ${\cal K}_1^{(1)}$ and ${\cal K}_3^{(1)}$, and obtain a consistent
covariant theory by solving for the remaining ${\cal K}$-functions. A
compatible metric perturbation $\delta\tilde{q}^{ab}$ can then also be derived.

The covariance condition therefore leaves room for modifications of
perturbative general relativity, even at the classical derivative order. This
result seems to confirm the standard treatment of loop quantum cosmology
because the background function ${\cal K}^{(0)}$ may well be chosen periodic
in $\bar{k}$ (or in $\bar{k}/\sqrt{|\bar{p}|}$ to model the Hubble
parameter). However, in contrast to a common assumption, the compatible
inverse spatial metric is not given by $|\bar{p}|^{-1}\delta^{ab}$, but rather by
\begin{equation}\label{eq:Background structure function - EMG}
    \bar{\tilde{q}}^{ab} = {\cal K}^{(1)}_3 {\cal K}^{(2)}_1 \frac{\delta^{ab}}{\bar p}
\end{equation}
where ${\cal K}_3^{(1)}$ is one of the free functions and ${\cal K}_1^{(2)}$
is determined by ${\cal K}^{(0)}$ and ${\cal K}_1^{(1)}$.

It is then possible to choose a modified covariant theory such that the
background dynamics is determined through some ${\cal K}^{(0)}$, and
independently require that the classical inverse spatial metric
$\tilde{\bar{q}}^{ab}=|\bar{p}|^{-1}\delta^{ab}$ is obtained. The equations
(\ref{eq:K^2_2=K^2_1 - simp})--(\ref{eq:K12 - simp}) then provide the full
dynamics of a compatible covariant theory for perturbative inhomogeneity on
the desired background. However, the resulting condition
${\cal K}_3^{(1)}{\cal K}_1^{(2)}=1$ implies a modified Hamiltonian
constraint whose perturbative terms are not guaranteed to follow the tenets of
loop quantum gravity. For instance, if ${\cal K}^{(0)}$ is a periodic function
in $\bar{k}/\sqrt{|p|}$, the most common choice in models of loop quantum
gravity, the derivative $\partial {\cal K}^{(0)}/\partial\bar{p}$ which
appears several times in (\ref{eq:K^2_2=K^2_1 - simp})--(\ref{eq:K12 - simp})
is not periodic in this variable or in $\bar{k}$. Moreover, solving these
equations may require dividing by periodic functions that do not fully cancel
out, in contrast to what is usually assumed in loop quantum cosmology.

Analogous constructions are available for perturbative gravity coupled to
scalar matter. However, the number of free functions increases to 58,
restricted by several complicated conditions which imply that the perturbation to the structure
function is given by \cite{EMGCosmoK,EMGCosmoD}
\begin{equation}\label{eq:Metric perturbation}
    \delta \tilde{q}^{ab} = \frac{1}{|\bar{p}|} {\cal K}^{(1)}_3 {\cal
      K}^{(2)}_1 \left( \frac{\delta^{aj} \delta E^b_j + \delta^{bj} \delta
        E^a_j}{|\bar{p}|} 
    - \delta^{a b} \frac{(\delta_d^k\delta E^d_k)}{|\bar{p}|}\right)\,.
\end{equation}
The perturbed inverse classical metric $\bar{q}^{ab}+\delta q^{ab}$
must therefore be multiplied by ${\cal K}^{(1)}_3 {\cal K}^{(2)}_1$ for covariant modifications.

\section{Singularities}
\label{s:Sing}

Previous proposals for the dynamics of perturbative inhomogeneity in loop
quantum cosmology did not consider complete covariance conditions and
therefore missed the correct form
\begin{equation}\label{eq:Inv metric EMG}
    \bar{\tilde{q}}^{ab} = {\cal K}^{(1)}_3 {\cal K}^{(2)}_1 \frac{\delta^{ab}}{|\bar p|}
\end{equation}
of the inverse background metric.  The proposal closest to a covariant
treatment previously available was made in \cite{ScalarHol}, see also
\cite{ScalarHolGen} for a recent extension. In this case, the brackets
(\ref{DD})--(\ref{HH}) were correctly implemented for perturbative
inhomogeneity, but the spatial metric was not adjusted to modifications with
${\cal K}^{(1)}_3 {\cal K}^{(2)}_1\not=1$. The resulting models are therefore
anomaly-free but non-covariant: While they define meaningful gauge theories
that provide the correct number of physical degrees of freedom, their
solutions for $\bar{q}_{ab}+\delta q_{ab}$ do not constitute the spatial part
of a well-defined space-time line element because their gauge transformations
are not compatible with coordinate changes. The corresponding cosmological
analysis, based on space-time concepts such as curvature and stress-energy,
geodesics, and horizons is therefore invalid.
This problem can easily be
corrected by evaluating resulting solutions of the equations of motion with
the emergent space-time metric
$\bar{\tilde{q}}_{ab}+\delta\tilde{q}_{ab}$. The additional factor of ${\cal
  K}^{(1)}_3 {\cal K}^{(2)}_1$ then has far-reaching implications for the
space-time geometry and its cosmological features.

\subsection{Basic ingredients}

For a specific analysis, we use the background modification
\begin{equation} \label{K0}
  {\cal K}^{(0)} = \frac{\sin^2(\lambda \bar{k})}{\lambda^2}
\end{equation}
in the gravitational constraint and keep all matter background terms in their
classical form (related to a scalar potential). In ${\cal K}^{(0)}$, $\lambda$
is a function parameterized as $\lambda(\bar{p})=\sqrt{\Delta} |\bar{p}|^\beta$
with two constants $\Delta>0$ and $\beta$. In \cite{ScalarHol}, the remaining
${\cal K}$-functions and one of the matter modifications are expressed in
terms of $\bar{k}$-dependent functions in a different way compared with what
we presented here, but as shown in \cite{EMGCosmoK} their solutions also solve
our consistency conditions. In addition to determining the remaining ${\cal
  K}$-functions beyond ${\cal K}^{(0)}$, the specific choices made in
\cite{ScalarHol} require $\beta=-1/2$, such that $\lambda=\sqrt{\Delta/|\bar{p}|}$.

The relevant functions that determine the emergent metric are now given by
${\cal K}_3^{(1)}=1$ and ${\cal K}_1^{(2)}=\cos(2\lambda\bar{k})$.  We
therefore obtain the inverse background metric
\begin{equation}
    \bar{\tilde{q}}^{ab} = \frac{\cos(2\lambda \bar{k})}{|\bar p|} \delta^{ab}
    \,.
\end{equation}
Since this expression is not guaranteed to be positive, it suggests signature
change in regions where $\cos(2\lambda \bar{k})<0$
\cite{Action,SigChange}. This observation has already been made for the model
given by \cite{ScalarHol}, but it was based on properties of the modified
constraint brackets as well as resulting equations of motion for propagating
scalar modes. Here, we complete the analysis by discussing implications of the
required $\bar{k}$-dependent factor in the emergent space-time line element.
As we will see, it will be sufficient to restrict attention to regions where
$\cos(2\lambda \bar{k})\geq0$, including the boundary. It is therefore not
required to discuss properties of mixed-type partial differential equations or
related conceptual questions that would be implied by signature change.

Our results, detailed in \cite{EMGCosmoK}, present both a covariantization as
well as a generalization of \cite{ScalarHol}. Previous results are made
covariant by replacing the classical inverse metric with the emergent one,
including an additional factor of $\cos(2\lambda \bar{k})$. Another
implication is that $\bar{k}$, at this level of modified gravity, only plays a
role of a canonical variable. It no longer directly represents extrinsic
curvature and therefore may well appear in the spatial metric. Curvature
(intrinsic or extrinsic) can then be derived in the standard way from the
emergent line element.  At the same time, the results generalize the previous
version by a systematic analysis of all relevant terms to the same order,
which leads to additional freedom in the choices of modification functions. In
particular, it is possible to find consistent covariant equations with
${\cal K}^{(0)}$ of the form (\ref{K0}) even if $\beta\not=-1/2$. The
lattice-refinement scheme \cite{InhomLattice} of loop quantum cosmology is
therefore not restricted by relevant consistency conditions.

\subsection{Effective Friedmann equation}

We return to the specific ${\cal K}$-modifications that
correspond to \cite{ScalarHol}.
The compatible background space-time line element is given by
\begin{equation} \label{dsEm}
  {\rm d}s^2=-\bar{N}{\rm d} t^2 + \frac{|\bar{p}|}{\cos(2\lambda\bar{k})}
  \delta_{ab} {\rm d}x^a {\rm d}x^b 
\end{equation}
with potential divergences implied by the new factor of
$1/\cos(2\lambda\bar{k})$. In order to analyze geometrical implications, we
need dynamical solutions for $\bar{p}$ and $\bar{k}$ as functions of
coordinate time. As a characteristic example, we use the modified background
Hamiltonian coupled to a free massless scalar field with a minimal
modification of coefficients. This, too, constitutes a standard choice in
models of loop quantum cosmology except for one of the perturbative momentum terms; see
\cite{EMGCosmoK} for details.

The background Hamiltonian constraint now equals
\begin{eqnarray} \label{Hpi}
    \tilde{H}^{(0)}[\bar{N}] &=& \int_{\Sigma}\mathrm{d}^3x\; \bar{N} \left( -
                                 \frac{3}{8\pi G} \sqrt{\bar p}
                                 \frac{\sin^2(\lambda(\bar{p}) \bar{k})}{\lambda(\bar{p})^2} +
                                 \frac{\bar{\pi}^2}{2\bar{p}^{3/2}} \right) 
\end{eqnarray}
where we assume $\bar{p}>0$ without loss of generality and, to be specific,
$\lambda(\bar{p})=\sqrt{\Delta/\bar{p}}$ or $\beta=-1/2$.  This expression
generates the background equations of motion
\begin{eqnarray}\label{eq:dot k - LQC}
    \dot{\bar k} &=& \frac{\bar{N}}{2\sqrt{\bar p}} \Bigg( \frac{\sin^2(\lambda\bar{k})}{\lambda^2}
    + 2 \bar{k} \frac{\sin(2\lambda\bar{k})}{2\lambda}
    - 8\pi G \frac{\bar{\pi}^2}{2\bar{p}^2} \Bigg)
    \,,\\
    \dot{\bar{p}} &=&
    2 \bar{N} \sqrt{\bar p} \frac{\sin(2\lambda\bar{k})}{2\lambda}
    \,,\label{eq:dot p - LQC}\\
    \dot{\bar\varphi} &=& \bar{N} \frac{\bar{\pi}}{\bar{p}^{3/2}}
    \,,\label{eq:dot phi - LQC}\\
    \dot{\bar\pi} &=& 0
    \,.\label{eq:dot pi - LQC}
\end{eqnarray}
Solving the constraint equation $\tilde{\cal H}^{(0)}=0$, we obtain
\begin{equation}\label{eq:H^0=0 - LQC}
    \frac{\sin^2(\lambda\bar{k})}{\lambda^2} = \frac{4\pi G}{3} \frac{\bar{\pi}^2}{ \bar{p}^2}
    =: \frac{8\pi G}{3} \bar{p} \bar{\rho}
    \,.
\end{equation}
For the specific form of $\lambda(\bar{p})$,  this equation implies
\begin{equation}\label{eq:p3 - holonomy}
    \bar{p}^3 = \frac{4\pi G}{3} \bar{\pi}^2 \frac{\Delta}{\sin^2(\lambda\bar{k})}
    \,,
\end{equation}
and hence the density parameter
\begin{equation}\label{eq:rho - holonomy}
    \bar{\rho} = \frac{\bar{\pi}^2}{2\bar{p}^3}= \frac{3}{8\pi G} \frac{\sin^2(\lambda\bar{k})}{\Delta}\,.
  \end{equation}

  Equation~(\ref{eq:dot p - LQC}) can then be written as
\begin{equation}\label{eq:Friedmann - LQC}
    \left(\frac{\dot{\bar{p}}}{2\bar{N}\bar{p}}\right)^2 =  \frac{8\pi G}{3} \bar{\rho}
                       \left(1-\frac{\bar{\rho}}{\rho_{\rm Q}}\right)
                     \end{equation}
                     where
                     \begin{equation}
    \rho_{\rm Q} = \frac{3}{8\pi G\Delta}
\end{equation}
is equivalent to our previous expression (\ref{rhoQell}) in terms of $\ell$
upon using the relationship between $p_q$ and $\bar{k}/\sqrt{\bar{p}}$. As
before, note that neither $\bar{k}/\sqrt{\bar{p}}$ nor
$\dot{\bar{p}}/(2\bar{p})$ can be identified with the observable Hubble
parameter because these expressions do not take into account the correct
spatial metric. Nevertheless, equation~(\ref{eq:Friedmann - LQC}) for the rate
of change $\dot{\bar{p}}/(2\bar{p})$ suggests that the trajectory
$\bar{p}(\tau)$ has a local minimum when
$\bar{\rho}=\rho_{\rm Q}$. At this time, we have
\begin{equation}
   \bar{p}= \bar{p}_{\Delta} = \left(\frac{4\pi G\bar{\pi}^2\Delta}{3}\right)^{1/3}
\end{equation}
and
\begin{equation}
    \lambda(\bar{p}) \bar{k}=\pi/2\,.
  \end{equation}

In the traditional interpretation of loop quantum cosmology with
(\ref{eq:Friedmann - LQC}) as an effective Friedmann equation, the local
minimum of $\bar{p}(\tau)$ is identified as a cosmological bounce \cite{APSII}. However,
this conclusion is based on the assumption that $\sqrt{\bar{p}}$ remains a valid
scale factor even after modifying the Hamiltonian constraint. Emergent
modified gravity shows that the correct geometrical scale factor, read off
from the emergent space-time line element (\ref{dsEm}), is instead given by
\begin{equation} \label{aem}
  \tilde{a}=\sqrt{\frac{\bar{p}}{\cos(2\lambda\bar{k})}}
\end{equation}
in terms of $\bar{p}$ and $\bar{k}$ as phase-space variables. (It is possible
to apply a canonical transformation that transforms $\bar{p}$ to $\tilde{a}^2$
and changes $\bar{k}$ to a new momentum. However, this transformation also
changes the modification functions and the diffeomorphism constraint in such a way that physically equivalent
results are obtained.)  Since $\bar{k}$ as well as $\lambda(\bar{p})$ are, in
general, time dependent, the cosine term contributes to the time derivative of
$\tilde{a}$ and to the Hubble parameter. The Friedmann equation then receives
further modifications not taken into account in previous discussions of loop
quantum cosmology.

Using the constraint, equation (\ref{eq:dot k - LQC}) can be written as
\begin{equation} \label{lkdot}
    (\lambda\bar{k})^\bullet = - \lambda \frac{\bar{N}}{2\sqrt{\bar p}} \left( 3 \frac{\sin^2(\lambda\bar{k})}{\lambda^2}
    + 4\pi G\frac{\bar{\pi}^2}{\bar{p}^2}\right)
    =
    - 4\pi G\bar{\pi}^2 \bar{N} \frac{\sqrt{\Delta}}{\bar{p}^3}\,.
\end{equation}
Combining this equation with $\dot{\bar{p}}$ and the constraint in order to
write trigonometric functions of $\lambda\bar{k}$ in terms of the 
density parameter $\bar{\rho}$, we obtain the new Friedmann equation
\begin{equation} \label{NewFriedmann}
\tilde{\cal H}^2=  \left(\frac{\dot{\tilde{a}}}{\bar{N}\tilde{a}}\right)^2 =
\frac{8\pi G}{3} \bar{\rho} 
                       \left(1-\frac{\bar{\rho}}{\rho_{\rm Q}}\right)
                       \frac{\left(1-8\bar{\rho}/\rho_{\rm
                               Q}\right)^2}{\left(1-2\bar{\rho}/\rho_{\rm
                               Q}\right)^2} \,.
\end{equation}
As our subsequent analysis will show, the new factor
  $\left(1-8\bar{\rho}/\rho_{\rm Q}\right)^2/\left(1-2\bar{\rho}/\rho_{\rm
      Q}\right)^2$ completely determines characteristic stages of the
  early-universe dynamics, while the old factor of
  $1-\bar{\rho}/\rho_{\rm Q}$, which is responsible for the traditional
  bounce in loop quantum cosmology, implies only minor modifications of the
  Hubble rate in cosmologically accessible regimes.
                           
If we approach the big bang in backwards evolution, starting with small 
density parameter $\bar{\rho}\ll\rho_{\rm Q}$, the first significant modification of
classical dynamics is reached when $\bar{\rho}=\frac{1}{8}\rho_{\rm Q}$, where the
emergent Hubble parameter $\tilde{\cal H}$ vanishes. This value implies a
local minimum of $\tilde{a}$, or a bounce. However, it is reached at a
density parameter less than $\rho_{\rm Q}$, and the dynamics around this minimum
is different from the traditional bounce because of the squares and extra
factors in (\ref{NewFriedmann}). Interestingly, and in contrast to
  traditional thinking in loop quantum cosmology, the bounce is not obtained
  at the maximum of the gravitational contribution to the Hamiltonian
  constraint (\ref{Hpi}) but rather at a smaller density parameter.

After the new bounce, the emergent scale factor as well as Hubble parameter
$|\tilde{\cal H}|$ increase in backward evolution until the next crucial phase
is reached when $\bar{\rho}=\frac{1}{2}\rho_{\rm Q}$. At this point, the
Hubble parameter diverges, implying a physical singularity as we will show
next. The putative bounce from traditional loop quantum cosmology, which
requires a density parameter of $\bar{\rho}=\rho_{\rm Q}$, is never reached in the
emergent geometrical picture. It lies beyond the physical singularity, in a
region where the line element (\ref{dsEm}) has Euclidean signature.

\subsection{Space-time analysis}

The analysis of singularities requires specific time-dependent solutions of
$\bar{p}$ and $\bar{k}$. The former can conveniently be obtained by starting
with (\ref{lkdot}) and choosing the time variable
\begin{eqnarray}
    T_K = \lambda\bar{k}/\sqrt{\Delta}\,.
\end{eqnarray}
This choice requires
$\dot{T}_K=1$, such that (\ref{lkdot}) together with
$\tilde{H}^{(0)}[\bar{N}]=0$ determines the background lapse function
\begin{equation}
    \bar{N} = -\frac{\bar{p}^3}{4\pi G\bar{\pi}^2}  = -
    \frac{\Delta}{3\sin^2(\sqrt{\Delta}\;T_K)}\,. 
\end{equation}

These solutions imply the space-time line element
\begin{eqnarray}\label{eq:Metric - internal time - LQC}
    {\rm d} s^2 &=& - \frac{\Delta^2}{\sin^4(\sqrt{\Delta} T_K)} \frac{{\rm d} T_K^2}{9}
    + \left(\frac{4\pi G \bar{\pi}^2\Delta}{3\sin^2(\sqrt{\Delta}
                    T_K)}\right)^{1/3} \frac{{\rm d} x^2+{\rm d}y^2+{\rm d}z^2}{\cos (2\sqrt{\Delta} T_K)}\,.
\end{eqnarray}
Proper time $\tau$ is related to $T_K$ by
$\cot(\sqrt{\Delta}T_K)=3\tau/\sqrt{\Delta}$, such that the remaining
functions in the spatial metric take the form
$\sin^2(\sqrt{\Delta}T_K)=1/(9\tau^2/\Delta+1)$ and $\cos(2\sqrt{\Delta}T_K)=
(9\tau^2/\Delta-1)/(9\tau^2/\Delta+1)$. 

Working in the original time coordinate $T_K$, the classical limit is reached at $T_K\to0$. Up
to a constant coefficient, the
geometrical scale factor determined by this line element is given by
\begin{equation}
  \tilde{a}(T_K)= \frac{\Delta^{1/6}}{ \sin(\sqrt{\Delta}
                    T_K)^{1/3}  \cos (2\sqrt{\Delta}
                      T_K)^{1/2}}
                \end{equation}
which is real in the range $0\leq \sqrt{\Delta}T_K\leq \pi/4$. Unlike the
classical behavior, reproduced for $\Delta\to0$, this function is not
monotonic but diverges at $T_K\to0$ (the classical asymptotics of infinite
expansion) and at $\sqrt{\Delta}T_k=\pi/4$. It reaches a non-zero minimum when
$\sin(\sqrt{\Delta}T_K)=1/(2\sqrt{2})$, halfway between the two divergences
in terms of the time coordinate $T_K$. In terms of proper time, the divergences are at $\tau\to\infty$ and $\tau=\frac{1}{3}\sqrt{\Delta}$, while
the minimum is obtained at $\tau=\frac{1}{3}\sqrt{7\Delta}$. The solution therefore implies a
bounce, but not at the value where $\dot{\bar{p}}$ vanishes and at a
density parameter smaller than the traditional value of $\rho_{\rm Q}$, given by
\begin{equation}
    \bar{\rho}_{\rm b} = \frac{\rho_{\rm Q}}{8}
  \end{equation}
  as expected by our analysis of the effective Friedmann equation (\ref{NewFriedmann}).

However, in spite of a bouncing background solution, we have an example of
singular cosmology. The line element has the Ricci scalar
\begin{eqnarray}
    R(T_K) &=& 2\frac{\sin^2 \left(\sqrt{\Delta}
               T_K\right)}{\Delta}\Bigg[29-52 \sin\left(2\sqrt{\Delta}
               T_K\right)- 41\cos \left(2 \sqrt{\Delta} T_K\right) 
    \nonumber\\
    &&\qquad\qquad\qquad\quad
    +52 \tan\left(2\sqrt{\Delta} T_K\right)+ \frac{9}{ \cos\left(2\sqrt{\Delta} T_K\right)}
    \Bigg]
       \,. 
\end{eqnarray}
This function diverges when $\sqrt{\Delta} T_K\to\pi/4$, or where
$\cos(2\lambda\bar{k})=0$, a time between the new bounce predicted by
(\ref{NewFriedmann}) and the traditional bounce of loop quantum
cosmology. Since the lapse function in (\ref{eq:Metric - internal time - LQC})
remains finite and non-zero at this value of our time coordinate, the
curvature divergence is reached at a finite value of proper time.  Amended by
using the compatible line element required for a covariant theory, the
modified equations given in \cite{ScalarHol} therefore imply a physical
singularity in the background geometry, even though neither $\bar{p}$ nor
$\tilde{a}$ reach the value zero. This model is not subject to dynamical
signature change because the Euclidean region is disconnected from the
Lorentzian region by the presence of a physical singularity.

\begin{figure}[h]
    \centering
    \includegraphics[clip=true,width=0.55\columnwidth]{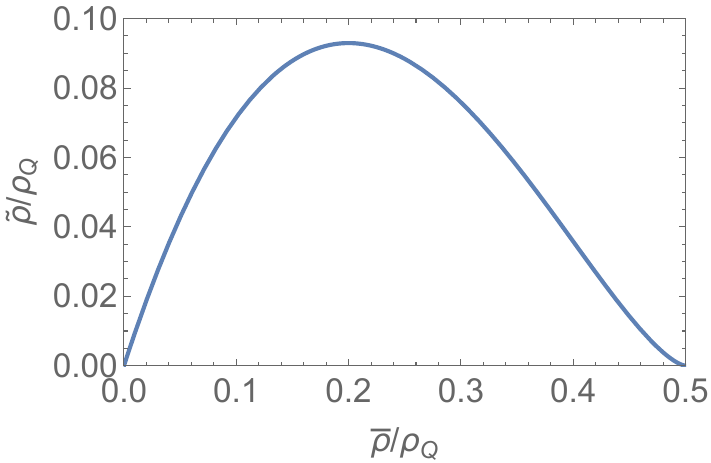}
    \caption{Emergent energy density $\tilde{\rho}$ as a function of the
      density parameter $\bar{\rho}$. The maximum
      $\bar{\rho}=\frac{1}{2}\rho_{\rm Q}$ is reached at the initial rip singularity,
      where the emergent energy density approaches zero due to infinite
      expansion. The maximum emergent density is more than one order of
      magnitude below the Planck density $\rho_{\rm Q}$.
    \label{fig:Emrho}}
\end{figure}

As a crucial new feature, instead of the usual big-bang singularity we
  have a (time-reversed) big-rip-type singularity \cite{phantom,Future,BigRip} at
  infinite scale factor. The emergent Hubble parameter is negative and
  diverges at the singularity, which constitutes the beginning of a collapse
  phase of finite duration that precedes a non-singular bounce. The density
  parameter $\bar{\rho}=\frac{1}{2}\rho_{\rm Q}$ is finite and reaches its
  maximum at the rip. It is non-zero in spite of the infinite expansion of the
  scale factor.

  However,
  the density parameter used so far
  differs from the emergent energy density that takes into account the derived
  space-time geometry: Using the emergent scale factor (\ref{aem}), 
  a spatial region with coordinate volume $V_0$ has the geometrical volume
  $V_0\tilde{a}^3$ and
  contains a scalar energy given by the matter contribution
  $H_{\pi}[1]=\frac{1}{2}V_0\bar{\pi}^2/\bar{p}^{3/2}$ to the Hamiltonian
  constraint (\ref{Hpi}). The latter has to be defined using the canonical variable
  $\bar{p}$ because the emergent scale factor is available only on-shell.
  The corresponding energy density should then be derived by dividing by the
  emergent volume. In this way, we obtain the emergent energy density
  \begin{equation}
    \tilde{\rho} = \frac{H_{\pi}[1]}{V_0\tilde{a}^3}= \bar{\rho}
    \cos^{3/2}(2\lambda\bar{k})= \bar{\rho} \left(1-\frac{16\pi G\Delta}{3}
      \bar{\rho}\right)^{3/2}\,. 
\end{equation}
This energy density is relevant for observational matter effects and physical
conditions that might be imposed on an initial state, while the density
parameter $\bar{\rho}$, as used initially, turns out to be more
convenient for a discussion of different cosmological eras. In
contrast to $\bar{\rho}$ and to classical examples of a big rip, the emergent energy density
$\tilde{\rho}$ approaches zero, see Fig.~\ref{fig:Emrho}.

    \begin{figure}[h]
    \centering
    \includegraphics[clip=true,width=0.55\columnwidth]{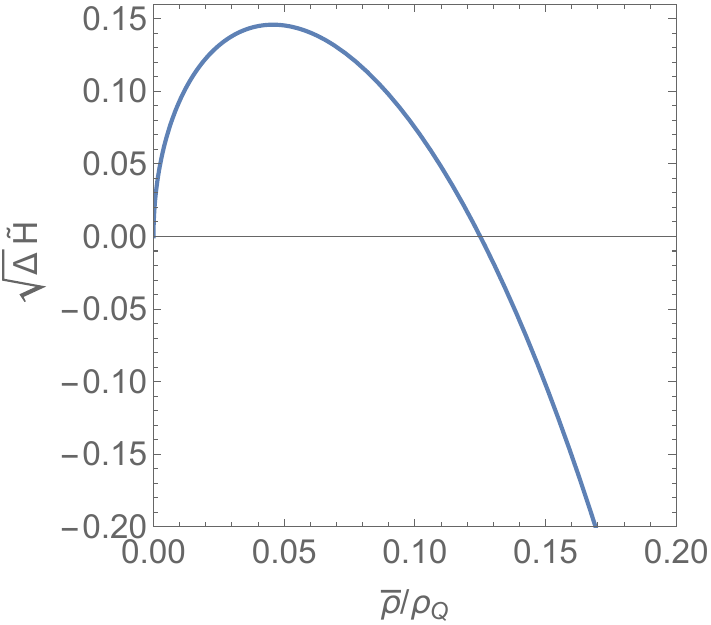}
    \caption{Rescaled emergent Hubble parameter $\sqrt{\Delta}\tilde{\cal H}$
      as a function of the rescaled density parameter $\bar{\rho}/\rho_Q$.}
    \label{fig:EmH}
\end{figure}

\begin{figure}[h]
    \centering
    \includegraphics[clip=true,width=0.55\columnwidth]{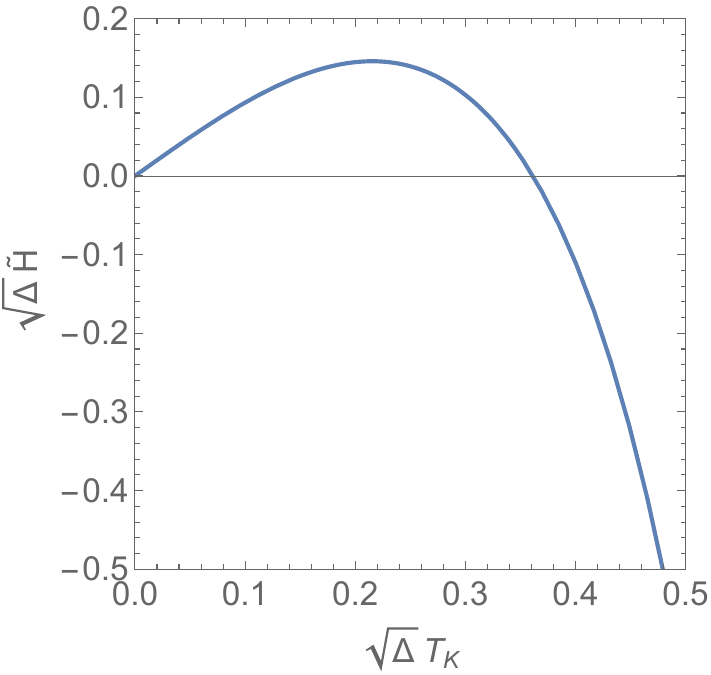}
    \caption{Rescaled emergent Hubble parameter $\sqrt{\Delta}\tilde{\cal H}$ as a function of the rescaled time coordinate $\sqrt{\Delta}T_K$.}
    \label{fig:EmHTK}
\end{figure}

\begin{figure}[h]
    \centering
    \includegraphics[clip=true,width=0.55\columnwidth]{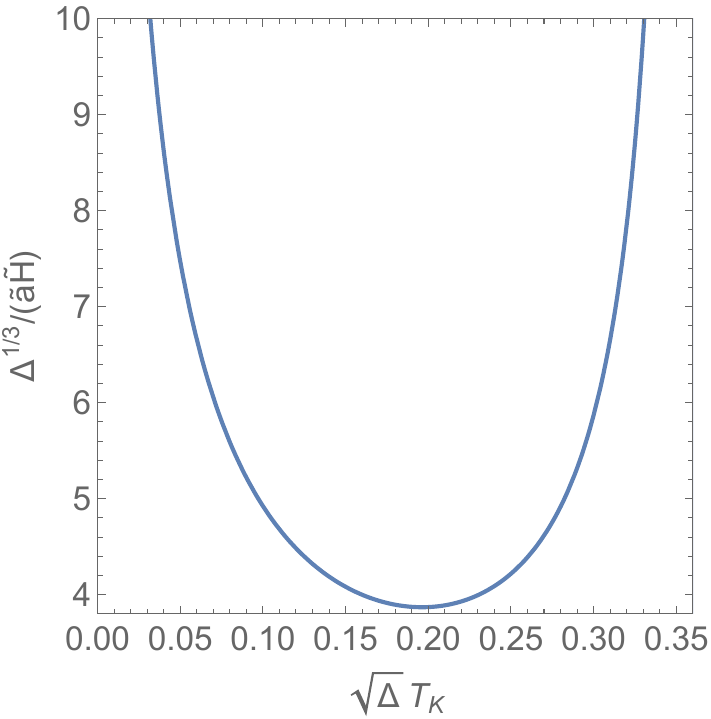}
    \caption{Rescaled comoving emergent Hubble radius after the bounce.}
    \label{fig:EmHr}
\end{figure}

While singular behavior in general complicates discussions of initial
values, a time-reversed rip can be expected to be more controlled than a big-bang
singularity. Our version is even tamer than the standard big rip because it
happens at vanishing emergent energy density.  The new dynamics may therefore
have advantages for cosmological model building along the lines of a
matter bounce \cite{MatterBounce}. In addition, after the bounce, the emergent
Hubble parameter $\tilde{\cal H}$ increases rapidly from zero at
$\bar{\rho}=\frac{1}{8}\rho_{\rm Q}$ to a maximum value at
$\bar{\rho}=\frac{1}{4} \left(2-6^{1/3}\right)\rho_{\rm
  Q}\approx0.046\rho_{\rm Q}$ or
$\sqrt{\Delta}T_K=\arcsin\left(1-6^{1/3}/2\right)\approx 0.22$; see
Figs.~\ref{fig:EmH} and \ref{fig:EmHTK}. The post-bounce comoving Hubble
radius, shown in Fig.~\ref{fig:EmHr}, has a minimum at
$\bar{\rho}\approx0.038\rho_Q$ obtained at a time
$\sqrt{\Delta}T_K\approx 0.20$, close to the maximum of the Hubble
parameter. In this density range, the emergent energy density $\tilde{\rho}$
is close to the density parameter $\bar{\rho}$.

  In contrast to the traditional bounce in loop quantum cosmology, all these
  energy scales are more than one order of magnitude below the Planck scale
  given by $\rho_{\rm Q}$. It is therefore reasonable to assume that the
  continuum geometry of emergent modified gravity can be used as a valid
  approximation of some underlying quantum theory of gravity in this
  regime. These scales are also closer to the typical density range in models
  of inflation.

\section{Non-singular model}
\label{s:NonSing}

As shown in \cite{EMGCosmoD}, an alternative covariant model exists where the
background modification (\ref{K0}) as well as
${\cal K}^{(2)}_1=\cos(2\lambda\bar{k})$ are preserved, but other modification
functions in the perturbed contributions to the constraint are changed,
including the function ${\cal K}^{(1)}_3=\sec(2\lambda\bar{k})$. This function
implies that the inverse background metric (\ref{eq:Inv metric EMG}) takes the
classical form
\begin{equation}
    \bar{\tilde{q}}^{ab} = \frac{\delta^{ab}}{|\bar p|}\,.
  \end{equation}
The compatible background space-time line element is then given by
\begin{equation}
  {\rm d}s^2=-\bar{N}{\rm d} t^2 + |\bar{p}|
  \delta_{ab} {\rm d}x^a {\rm d}x^b 
\end{equation}
  and the emergent scale factor equals $\sqrt{|\bar{p}|}$.

  Because the background Hamiltonian constraint is identical to that of the
  previous model, the same equations of motion and dynamical solutions are
  obtained for the background.  However, the different dependence of the line
  element on the phase-space variables implies a different space-time geometry, now
  described by
\begin{eqnarray}\label{eq:Metric - internal time - EMG}
    {\rm d} s^2 &=& - \frac{\Delta^2}{\sin^4(\sqrt{\Delta} T_K)} \frac{{\rm d} T_K^2}{9}
    + \left(\frac{4\pi G \bar{\pi}^2\Delta}{3\sin^2(\sqrt{\Delta} T_K)}\right)^{1/3} \left({\rm d} x^2+{\rm d}y^2+{\rm d}z^2\right)\,.
\end{eqnarray}
Unlike (\ref{eq:Metric - internal time - LQC}), this line element remains
regular at $\sqrt{\Delta} T_K=\pi/4$. Space-time therefore extends beyond this
hypersurface without encountering a singularity or suffering signature change.  The
Ricci scalar of this line element is given by
\begin{eqnarray}
    R = - 6 \frac{\sin^2 \left(\sqrt{\Delta} T_K\right)}{\Delta} \left(1-4\sin^2 \left( \sqrt{\Delta} T_K\right)\right)\,,
\end{eqnarray}
which is finite everywhere.

For this model, which is governed by the traditional effective Friedmann
equation, the minimum scale factor does occur when $\dot{\bar{p}}$ vanishes
and the energy density reaches the value of $\rho_Q$.  The divergences in the
line element (\ref{eq:Metric - internal time - EMG}) at $T_K=0$ and $T_K=\pi$,
correspond, respectively, to the classical asymptotics of infinite expansion
in the distant future $\tau\to\infty$ and collapse in the remote past
$\tau\to-\infty$.  Finally, while the two models share the same dynamics for
the background phase-space variables, though with different background
geometries, the difference in the modification functions of the perturbed
contribution to the constraints implies different perturbative dynamics.  This
and further physical implications are discussed in detail in
\cite{EMGCosmoD}.

In this model the Hubble parameter maximizes at $\rho_Q/2$ or
  $\sqrt{\Delta}T_K=\pi/4$, which could be used for matter bounce cosmology or
  a partial inflationary period at an elevated near-Planckian density
  scale. The background dynamics is identical with traditional loop quantum
  cosmology, but perturbative inhomogeneity is subject to new equations that
  implement a covariant space-time geometry without background
  singularities. However, some of the required coefficients in the equations
  for inhomogeneity, such as ${\cal K}^{(1)}_3=\sec(2\lambda\bar{k})$, diverge
  near the bounce.  Self-consistent dynamics then requires solutions in which
  inhomogeneity remains small through the bounce phase in which some of the
  background coefficients are large. Evolving perturbations through the bounce
  is therefore challenging in such a model, even in the absence of signature
  change compared with \cite{ScalarHol}. Further analysis of the perturbative
  dynamics is required in order to determine whether background rip-type
  singularities of the first model or divergent coefficients in the equations
  for inhomogeneity in the second model are more controllable for physical
  model building. The vanishing emergent energy density in the former case,
  combined with the fact that all phase-space variables remain regular,
  might suggest a preference for this version.

\section{Conclusions}

We have revisited constructions of anomaly-free models of cosmological
perturbations suitable for some effects expected from loop quantum cosmology,
as first introduced in \cite{ScalarHol}. Previous equations were consistent
gauge theories but did not incorporate full covariance conditions. The tensors
used to define space-time geometry were therefore incorrect. Emergent modified
gravity provides a suitable setting to complete the analysis and derive a
well-defined space-time geometry.

On a formal level, the analysis implies an apparently small amendment of the
equations found in \cite{ScalarHol}, given by an additional factor in the
spatial part of the space-time metric. This factor is not positive definite
and is therefore consistent with previous interpretations of modified
equations in terms of signature change. However, it had not been included in
the space-time line element in previous cosmological studies, and therefore
its most important geometrical consequence had been missed: It implies a
curvature singularity at the hypersurface that had formally been identified as
the place of signature change. Even though the underlying phase-space dynamics
may still appear to suggest the standard non-singular bounce of loop quantum
cosmology, the compatible space-time geometry for background and perturbations
has a physical singularity at the corresponding density. We derived a new
effective Friedmann equation (\ref{NewFriedmann}) which implies a bounce at a
smaller energy density, preceded by a collapsing phase that starts at a
time-reversed rip-type singularity with infinite
scale factor. 

With this result, current models of loop quantum cosmology share the singular
beginning with standard big-bang cosmology, but with a tamed version of a rip
singularity. The new collapsing phase and the subsequent bounce crucially
change the early-universe behavior by including various regimes of a changing
Hubble radius at energy scales more than an order of magnitude below the
Planck scale.  Our complete covariance analysis, guided by emergent modified
gravity, was essential for seeing this conclusion.

The same broad framework also opens up options for the construction of new
covariant models that are free of physical singularities in the background
dynamics. The emergent metric coefficients depend on the modification
functions used in the Hamiltonian constraint, which are subject to a large
number of rather involved consistency conditions from anomaly freedom and
covariance discussed in detail in \cite{EMGCosmoK,EMGCosmoD}. The specific
form of the emergent line element may be used to find modifications that imply
non-singular space-time solutions, for which one example is given in
(\ref{eq:Metric - internal time - EMG}). The form of these modification
functions, which differ from the singular example given in \cite{ScalarHol},
then allow conclusions about specific quantization steps of loop quantum
gravity that could lead to such terms at an effective level. For instance, the
modification function ${\cal K}^{(1)}_3=\sec(2\lambda\bar{k})$ used for the
model of Section~\ref{s:NonSing}, which eliminates background singularities,
would not be expected by traditional loop constructions which rather make use
of polynomial expressions of trigonometric functions. Although a variety of
consistent models for perturbative inhomogeneity is possible in emergent
modified gravity, the conditions are strong enough to place interesting
constraints on possible constructions for models of loop quantum gravity.

\section*{Acknowledgements}

This work was supported in part by NSF grant PHY-2206591.


\end{document}